\documentstyle[aps,12pt]{revtex}
\begin{document}
\title{Phase Transitions in the Symmetric Kondo Lattice Model \\
in Two and Three Dimensions}
\author{Zhu-Pei Shi and Rajiv R. P. Singh}
\address{Department of Physics, University of California, Davis,
  California 95616}
\author{Martin P. Gelfand}
\address{Department of Physics, Colorado State University, Fort Collins,
  Colorado 80523}
\author{Ziqiang Wang}
\address{Department of Physics, Boston University, Boston, Massachusetts 02215}
\maketitle
\vskip 3cm
\centerline{\bf Abstract}
\begin{abstract}
We present an application of high-order series expansion in
the coupling constants for the ground state properties
of correlated lattice fermion systems.
Expansions have been generated up to order $(t/J)^{14}$ for
$d=1$ and $(t/J)^8$ for $d=2,\ 3$ for
certain properties of the symmetric Kondo lattice model.
Analyzing the susceptibility series, we find evidence for a continuous
phase transition from the ``spin liquid'' phase characteristic
of a ``Kondo Insulator'' to an antiferromagnetically ordered phase
in dimensions $d\ge2$ as the antiferromagnetic Kondo coupling is decreased.
The critical point is estimated to be at $(t/J)_c\approx0.7$ for
square lattice and $(t/J)_c\approx0.5$ for simple-cubic lattice.
\end{abstract}
 \begin{verbatim}
 PACS numbers: 75.30.Mb, 75.30.Kz, 75.20.Hr, 75.40.Mg
 \end{verbatim}
\eject
\narrowtext
The interplay between magnetic Ruderman-Kittel-Kasuya-Yoshida (RKKY)
interaction and the Kondo effect is a central
unresolved problem in the physics of valence fluctuation and
heavy fermion compounds \cite{Lee1}.
The development of magnetic long-range order in heavy fermion
metals has been a subject of intensive experimental investigation.
On the theoretical side,
aside from an early conjecture made by Doniach \cite{Doniach1} and Varma
\cite{Varma1} that a state with antiferromagnetic (AF)
long-range order is energetically
more favorable than the non-magnetic Kondo state for small enough AF Kondo
couplings, such basic issues as the location of the critical point, the
order of the transition, and the transport properties in the critical
regime remain to be understood \cite{Lee1}.
The recent revival of interest in the class of stoichiometric
insulating compounds known as ``Kondo
insulators'' \cite{Aeppli1} has uncovered {\it both} non-magnetic
semiconducting and AF-ordered materials
at low temperatures.
Examples of the former are CeNiSn, Ce$_3$Bi$_4$Pt$_3$, and CeRhSb
\cite{Aeppli1}, whereas those of the latter are
UNiSn \cite{Bykovetz}, CePdSn and CePtSn \cite{Kasaya}.
Furthermore, susceptibility and transport measurements on
ternary compounds CeNi$_{1-x}$(Pd,Pt)$_x$Sn
suggest a continuous AF transition as a function
of isoelectronic substitutions $x$ \cite{Kasaya}.
In this context, the periodic Anderson model and
the symmetric Kondo lattice model (KLM) may
be considered as prototypical models that plausibly captures some of
the essential physics of electronic correlations in stoichiometric
Kondo insulators \cite{Doniach2}.
The possibility of a continuous transition between an insulating,
spin-liquid phase and an AF phase
in these models has been investigated recently by
mean-field approximations \cite{Schlottmann},
variational wavefunction calculations \cite{Fazekas1,Wang1} and
large-scale quantum Monte Carlo simulations \cite{Vekic1}.

In this Communication we apply high-order Rayleigh-Schr\"odinger
perturbation theory \cite{Gelfand1},
about the strong-coupling limit, to the zero-temperature
properties of the symmetric KLM.
In particular, we address the possibility of a
continuous transition in dimensions $d\ge2$.
The Hamiltonian of the symmetric  KLM is given by
\begin{equation}
H=-t\sum_{\langle ij\rangle\sigma}c^\dagger_{i\sigma}c_{j\sigma}
+ J\sum_i \bbox{S}_i \cdot \bbox{s}_i,
\end{equation}
which describes a half-filled band of conduction electrons
(with creation operator $c_{i\sigma}^\dagger$ and nearest neighbor
hopping amplitude $t$)
interacting with a lattice of spin-half
local moments $\bbox{S}_i$ through an intra-site Kondo coupling
$J$ (taken to be AF, i.e. $J>0$);
$\bbox{s}_i={1\over2} c^\dagger_{i\sigma}
\bbox{\tau}_{\sigma\sigma'} c_{i\sigma'}$
(where $\bbox{\tau}$ are the Pauli matrices) denotes the conduction
electron spin at site $i$.
We will consider only the simplest lattice structures in this paper,
so that $d=2$, 3 is shorthand for the square and simple-cubic lattices,
respectively.

In one dimension, the symmetric KLM possesses an insulating spin-liquid ground
state characteristic of a ``Kondo insulator'' for all values of the
Kondo coupling $J>0$. The spin and charge excitations
exhibit interesting properties found
through quantum Monte Carlo (QMC) simulations \cite{Fye1,Troyer1},
exact diagonalization \cite{Tsunetsugu1}, variational
wavefunction \cite{Wang2}, and density matrix renormalization
group (DMRG) \cite{Yu1} studies.
For a $N$-site symmetric KLM, the number of states with
$\sum_i (S^z_i + s^z_i)=0$ (that is, the dimension of the
space within which finite-lattice calculations are generally done
in the absence of any spatial symmetries)
is given by $\sum_m {N\choose m}^3$,
a number which reaches roughly 740,000 for $N=8$.
Exact diagonalization is therefore limited to systems with $N\le 10$
\cite{Tsunetsugu1},
which suggests that approach is only suited to the study
of the one-dimensional KLM.
The DMRG, which has been applied to the symmetric KLM
with $N$ up to 24 \cite{Yu1}, is also effectively restricted
to one-dimensional (or quasi-one-dimensional) systems so far
\cite{Liang1}. The large Hilbert space makes the generalization
of these methods to $d>1$ impractical at the present time.

The series expansions, unlike other $T=0$ numerical
approaches, is not restricted
to $d=1$, and it is a method well-suited to locating
critical points (which there is reason to believe exist in $d\ge2$).
Its primary requirement is that the unperturbed Hamiltonian
have a nondegenerate (or finitely degenerate)
ground state, which is fulfilled by the symmetric KLM in the
strong-coupling ($t/J=0$) limit.
To demonstrate the validity of our series expansions,
the $d=1$ symmetric KLM will be first considered
and our results compared with others' in the literature.
For $d\ge2$, our calculations suggest that AF
spin susceptibilities for both local moments and conduction electrons
diverge at a critical value of $t/J$, indicative of a continuous
AF-ordering phase transition. Our estimates of the critical points are
$(t/J)_c\approx0.7$ for $d=2$ and $(t/J)_c\approx0.5$ for $d=3$.

In the strong coupling limit, the ground state is non-degenerate
and describes a state in which the conduction electron spin and
the local moment are locked into local singlets at every site.
The series calculations involve perturbative diagonalization
of finite-cluster Hamiltonians, with the hopping term ($t$) as the
perturbation,
followed by evaluation of relevant matrix elements.
In contrast with exact diagonalization, perturbative diagonalization
is a noniterative process: with $m$ matrix multiplications one obtains
the ground state eigenvector to order $(t/J)^m$ and the ground state
energy to order $(t/J)^{m+1}$.

Properties of the KLM for which series have been constructed include:
the ground state energy per site $E$;
the zero-frequency local moment and conduction electron antiferromagnetic
spin susceptibilities $\chi_l(\bbox{Q})$ and $\chi_c(\bbox{Q})$
(where $\bbox{Q}$ is the zone-corner wavevector, and it is assumed that
$g=2$ for both conduction and localized electrons);
equal-time two-point density correlations ${\cal N}(\bbox{r}) =
\langle (n(\bbox{r}) - 1)((n(\bbox{0}) - 1) \rangle$
(with $n=\sum_\sigma c^\dagger_\sigma c_\sigma$);
equal-time single-particle Green's functions
${\cal G}(\bbox{r}) =
\langle \sum_\sigma (c^\dagger_\sigma(\bbox{r})c_\sigma(\bbox{0}) +
c^\dagger_\sigma(\bbox{0})c_\sigma(\bbox{r})) \rangle$
and thence the momentum distribution function
$n(\bbox{k})=\sum_{\bbox{r}}{\cal G}(\bbox{r})\exp(i\bbox{k}\cdot\bbox{r})$;
and local moment correlations
${\cal S}_l(\bbox{r}) =
\langle S^z(\bbox{r}) S^z(\bbox{0}) \rangle$.
Above, $\langle\cdot\rangle$ denotes the ground-state
expectation value.

The cluster method (connected graph expansion)
is used to carry out the series expansion.
For a description of the cluster method in the
context of quantum spin systems see Ref.~\cite{Gelfand1}
(but note that substantial technical improvements have been
made in the weight-calculation algorithms since that writing).
The essence of the cluster method is that one can express
quantities such as those listed above in terms of sums over clusters
$\sum_g L(g) W(g,t/J)$, where $L$ is the ``lattice constant''
which describes the number of embeddings of the cluster into the
lattice and $W$, a power series in $t/J$, is the ``weight'' of the cluster $g$.
For all of the above properties, the clusters
one must consider are just the naively connected ones, in which
lattice sites correspond to nodes of a graph and the terms in
the kinetic energy correspond to edges.  For the one-dimensional
case, the clusters are simply open chains.
One knows on quite general grounds that the leading nonzero term in
$W$ for a cluster with $n$ edges is of order $(t/J)^{n}$.
Thus, by considering all distinct clusters
with up to $n$ sites one can determine the coefficients
in the expansion for some property to order $(t/J)^{n}$.
In the following presentation,
the results for physical quantities are expressed as
$J^\alpha \sum_n c_n(t/J)^n$, where $c_n$ are the series coefficients
while $\alpha=-1$ for the susceptibilities,
$\alpha=1$ for the energy $E$, and
$\alpha=0$ for other expectation values.

We note in passing that the series presented here have some special properties
which follow from particle-hole symmetry of the symmetric KLM
on a bipartite lattice.  Most of the series contain only even powers
of $t/J$; most graphs (for most properties) have leading nonzero terms
in their weights of higher order than $(t/J)^n$.  In consequence, although
we are presently limited to calculations for 8-site clusters, we know
the exact expansions to order $(t/J)^{14}$ in $d=1$;
and exact expansions to order
$(t/J)^8$ in $d=2$ (3) are obtained by consideration of only 17 (18)
topologically distinct clusters of which only 1 (2) has 8 sites.

Next we present the results, beginning with $d=1$.
The first test of the series comes from an examination
of the ground-state energy; the coefficients are listed
in Table~\ref{table:1d}.  To obtain estimates of $E$ at finite $t$,
ordinary Pad\'e approximants are employed; a comparison
of the best-behaved ($[4/3]$) approximant with the result from
the DMRG calculation for
a 24-site chain is displayed in Fig.~\ref{fig:1}.
The agreement is excellent up to $t/J\sim0.75$.
One can also compare ${\cal S}_l$ with the DMRG results;
at $t/J=0.5$, series estimates of the nearest and
second-neighbor correlations differ from
DMRG estimates by only 0.2\%.
We have calculated the conduction electron
momentum distribution $n(k)$ for different values of $t/J$. The results
shown in Fig.~\ref{fig:n_of_k} are typical for an insulator.
The results for $t/J=0.625$ compare well with those obtained
using quantum Monte Carlo for systems with up to eight sites Ref.~\cite{Fye1}.
To illustrate what one gains by
going to the effort of calculating a high-order series,
comparisons are provided in Table~\ref{table:compare} to
the results of lowest-order perturbation theory and other
numerical calculations.

With the results for one dimensional systems validating the
accuracy of the series calculations, we now turn to $d\ge2$.
The relevant physical distinction between $d=1$ and $d\ge2$
is that an $S=1/2$ antiferromagnet orders at $T=0$ only for $d\ge2$.
The question at hand, as discussed in the introduction, is
whether there is evidence for a transition to an
AF-ordered phase as $t/J$ increases from zero, and, if so, what
is the value of $t/J$ at the transition. Let us first consider $d=2$.
Series for $E$, the local-spin structure
factor $\hat{\cal S}_l(\bbox{Q})$ (the Fourier transform of
${\cal S}_l$), $\chi_l(\bbox{Q})$ and $\chi_c(\bbox{Q})$
are listed in Table~\ref{table:2d}.

The AF susceptibility series have
positive, monotonically increasing terms and
are thus strongly suggestive of a critical point at some $(t/J)_c<1$
at which the AF correlations become long-ranged.
The ratios of consecutive terms $c_n/c_{n-2}$ do not vary smoothly
with $n$, so from these rather short series it is
difficult to obtain a precise estimate of $(t/J)_c$.
Analysis of the $\chi_l$ series by means of
differential approximants \cite{ida}
yields no reasonable approximants.
For the $\chi_c$ series, there are three plausible approximants
with singularities at $(t/J)^2=0.60$, 0.84, and 0.48; and
there is a correlation between the locations of the singularities
and the values of the associated exponents.
If the charge excitations remain gapful at this magnetic
transition (as naively expected, and also supported by the
variational calculation \cite{Wang1}),
the transition should lie in the $d=3$ classical
Heisenberg model universality class with known exponent $\gamma\approx1.4$.
Interpolating to this value of $\gamma$
yields an estimate for $(t/J)_c^2$
of 0.53, with an uncertainty of 0.02 associated with the
interpolation alone.

Because the series are short, it is worthwhile to
also examine the {\it direct\/} Pad\'e approximants to the series.
Doing so is tantamount to biasing $\gamma$ to
the value 1; since the correct value is not too much different
the locations of the poles might not be too far off, for our purposes.
(In fact, one can apply this method to first five or six terms of
the classical, simple cubic lattice Ising model susceptibility
high-temperature expansion, and find
that consistent estimates of the critical coupling are obtained which
are only 10\% too small.)
The $[2/2]$ approximants for the series yield poles at $(t/J)^2=0.44$
and 0.69 for the conduction and local electron susceptibilities, respectively.

The bottom line is that out best estimate for the
AF ordering critical point in $d=2$
comes from consideration of $\chi_c(\bf{Q})$:
$(t/J)_c \approx 0.7$, with an uncertainty of roughly 0.1.
This is in excellent agreement with the
variational calculation of Wang {\it et al.}~\cite{Wang1,Wang2},
but differs from the estimate (a value near $2.1$)
of Fazekas and M\"uller-Hartmann ~\cite{Fazekas1}.
The underlying reason for that agreement with Wang {\it et al.} is not obvious;
let us just note that that class of variational wavefunctions
becomes exact in the strong-coupling limit, which is also the starting
point of the series expansion.

An analogous analysis has been carried out for the $d=3$ susceptibility
series listed in Table~\ref{table:3d}.  In this case it is appropriate
to bias the critical exponent to ``1 plus logarithmic corrections'' which
we take to mean somewhere in the range 1 to 1.1.  The differential
approximants are more consistent than in $d=2$: the plausible approximants
for $\chi_c(\bbox{Q})$ have poles at 0.235, 0.256 and 0.281 with
corresponding exponents 1.02, 1.36, and 1.51.
The same is true for the ordinary Pad\'es: the [2/2]
approximants for $\chi_c(\bbox{Q})$ and $\chi_l(\bbox{Q})$ have poles
at $(t/J)^2=0.23$ and 0.22, respectively.  Our preferred estimate
is $(t/J)_c=0.49$ with a conservative uncertainty of 0.04.

In summary, a cluster expansion technique for generating high-order
$T=0$ perturbation expansions for quantum many-body systems has
been successfully applied to the symmetric KLM in $d=1$, 2, and 3.
For $d=1$, physical quantities estimated
by appropriately extrapolated perturbation expansions are reliable
up to intermediate values of $t/J$.
In $d\ge2$, we have obtained
estimates of the critical coupling separating the
Kondo-insulating, spin-liquid phase
at small $t/J$ from an antiferromagnetically ordered phase;
the transition results from competition between the Kondo
effect and RKKY interactions.
This property of the symmetric KLM may be relevant to
the observed AF compounds and the transition from nonmagnetic insulator
to antiferromagnet in the class of heavy fermion insulators
\cite{Bykovetz,Kasaya}.
The properties of the charge excitations across this transition
is an interesting issue that remains to be addressed.

{\it Acknowledgments.} MPG is supported by NSF Grant DMR 94--57928;
he is grateful for the hospitality of Kevin Bedell during a
visit to Los Alamos National Laboratory which was supported by
Associated Western Universities.
RRPS and ZS are supported by NSF Grant DMR 93--18537; ZS received further
support from the University Research fund of the University of California,
Davis.  We are grateful to Clare Yu and Richard Fye for sharing the detailed
numerical results of their calculations, and to Barry Klein
and Dung-Hai Lee for discussions.

\vskip 2cm
\centerline{\bf Figure Captions}
\begin{figure}
\caption{Ground state energy $E$ for the one-dimensional
symmetric KLM, comparing the best Pad\'e approximant (solid line) to the
series with the nearly exact results from the
DMRG (filled circles with a dashed line as guide to the eye)
\protect\cite{Yu1}.}
\label{fig:1}
\end{figure}

\begin{figure}
\caption{Momentum distribution $n(k)$ for the one-di\-men\-sional
symmetric KLM for several values of $t/J$, as determined
from the series expansions for ${\cal G}({\bf r})$.}
\label{fig:n_of_k}
\end{figure}

\eject
\centerline{\bf Tables}
\medskip


\narrowtext
\begin{table}
\caption{Series coefficients $c_n$ for the energy and first and
second neighbor local spin correlations
of the one-dimensional KLM.  Note that the coefficients of odd powers
vanish for these properties.}
\begin{tabular}{cccc}
$n$ & $E$ & ${\cal S}(1)$ & ${\cal S}(2)$ \\
\hline
0  & $-$0.75 & 0.0 & 0.0 \\
2  & $-$0.6666666667 & $-$0.2777777778 &   0.0\\
4  & $-$0.3111111111 &  0.0111111111 &   0.2911111111 \\
6  &  0.7114685479 &  1.0581324823 &  $-$0.3128030682 \\
8  &  0.1925692156 & $-$1.2088861341 &  $-1$.2886446333 \\
10 & $-$2.8528410569 & $-$5.5271656257 &   3.7729247608 \\
12 &  2.2809484235 & 16.5201887345 &   4.1162298390 \\
14 & 12.8828505218 & 19.6847763427 & $-$35.1869342747 \\
\end{tabular}
\label{table:1d}
\end{table}

\mediumtext
\begin{table}
\caption{Comparison of lowest nontrivial order in perturbation theory
(PT) (second order, except for ${\cal S}_l(2)$ for which it is fourth);
extrapolated high-order perturbation theory (with uncertainties of 1
in the last digit, as estimated by the consistency of various Pad\'e
approximants); and
other calculations of selected properties of the one-dimensional KLM.}
\begin{tabular}{dcddd}
 $t/J$ & Property & Lowest-order PT & Extrapolated PT & Other calculations \\
\hline
 0.5   &  $E$  & $-$0.9167  &  $-$0.9261 & $-$0.9261\tablenotemark[1] \\
 0.5   & ${\cal S}_l(1)$ & $-$0.0695 & $-$0.0587 & $-$0.05875
 \tablenotemark[2]\\
 0.625 & ${\cal S}_l(1)$ & $-$0.1085 & $-$0.0788 & $-$0.088(10)
 \tablenotemark[3] \\
 0.5   & ${\cal S}_l(2)$ &    0.01819 & 0.01157  & 0.01155
 \tablenotemark[2]\\
 0.625 & ${\cal S}_l(2)$ &    0.04442 & 0.0212   &
 0.021(5)\tablenotemark[3]  \\
\end{tabular}
\tablenotetext[1]{Obtained by extrapolating, versus $1/N$, the DMRG\cite{Yu1}
results for open chains consisting of $N=16$ and 24 sites.  Note that
for an open 24-site ring $E=-0.919159$, so extrapolation is
necessary to obtain an accurate estimate of $E$ from open-chain calculations.}
\tablenotetext[2]{DMRG for 24-site open chain.\cite{Yu1}}
\tablenotetext[3]{Quantum Monte Carlo for closed 8-site rings, with
boundary-condition averaging.\cite{Fye2}}
\label{table:compare}
\end{table}

\narrowtext
\begin{table}
\caption{Series coefficients $c_n$ for the antiferromagnetic
local-spin structure factor, and the zero-frequency antiferromagnetic
spin susceptibilities for the localized and conduction electrons,
in the square-lattice KLM. The coefficients for odd $n$ all vanish.}
\begin{tabular}{cccc}
$n$ & $\hat{\cal S}_l(\bbox{Q})$ & $\chi_l(\bbox{Q})$ & $\chi_c(\bbox{Q})$ \\
\hline
0  &  $-$0.25       &   2.0        &   2.0 \\
2  &     1.55555556 &  26.07407407 &   7.11111111 \\
4  &     1.43506173 & 154.03667490 &  18.53155556 \\
6  & $-16$.90670160 & 162.70126205 &  24.94787492 \\
8  &   105.05854716 & 267.49655192 & 173.60536516
\end{tabular}
\label{table:2d}
\end{table}

\narrowtext

\begin{table}
\caption{
Same as in Table III, but for simple-cubic-lattice KLM}
\begin{tabular}{cdd}
$n$ &  $\chi_l(\bbox{Q})$ & $\chi_c(\bbox{Q})$ \\
\hline
 0  &     2.0       &    2.0 \\
 2  &    39.1111111 &   10.6666667 \\
 4  &   326.9517696 &   49.2918518 \\
 6  &   746.1135610 &  171.9991437 \\
 8  &  6739.6417011 & 1493.2844373
\end{tabular}
\label{table:3d}
\end{table}

\end{document}